\documentclass[12pt,a4paper]{article}
\usepackage{amsmath}
\usepackage{braket}
\usepackage{amsfonts}
\usepackage{amssymb}
\usepackage{graphicx}
\usepackage{amsthm}
\usepackage{listings}
\usepackage{float}
\usepackage{caption}
\usepackage{subcaption}
\usepackage{color}
\usepackage[colorlinks=true,allcolors=blue]{hyperref}
\usepackage{chngcntr}
\usepackage{cite}
\usepackage{authblk}
\usepackage{bm}
\counterwithout{figure}{section}
\usepackage[left=2.00cm, right=2.00cm, top=2.00cm, bottom=2.00cm]{geometry}
\title{Coherent states for graphene under the interaction of crossed electric and magnetic fields}
\date{ }
\author[1]{M Castillo-Celeita\footnote{mfcastillo@fis.cinvestav.mx}}
\author[1,2]{E D\'iaz-Bautista\footnote{ediaz@fis.cinvestav.mx, ediazba@ipn.mx}}
\author[1]{M Oliva-Leyva\footnote{mauriceoliva.cu@gmail.com}}
\affil[1]{\small Physics Department, Cinvestav, P.O. Box. 14-740, 07000 Mexico City, Mexico}
\affil[2]{\small Departamento de Formaci\'on B\'asica Disciplinaria, Unidad Profesional Interdisciplinaria de Ingenier\'ia Campus Hidalgo del Instituto Polit\'ecnico Nacional, Pachuca: Ciudad del Conocimiento y la Cultura, Carretera Pachuca-Actopan km 1+500, San Agust\'in Tlaxiaca, 42162 Hidalgo, Mexico}
\begin{document}
	\maketitle
	
	\begin{abstract}
We construct the coherent states for charge carriers in a graphene layer immersed in crossed external electric and magnetic fields. For that purpose, we solve the Dirac-Weyl equation in a Landau-like gauge avoiding applying techniques of special relativity, and thus we identify the appropriate rising and lowering operators associated to the system. We explicitly construct the coherent states as eigenstates of a matrix annihilation operator with complex eigenvalues. In order to describe the effects of both fields on these states, we obtain the probability and current densities, the Heisenberg uncertainty relation and the mean energy as functions of the parameter $\beta=c\,\mathcal{E}/(v_{\rm F}B)$. In particular, these quantities are investigated for magnetic and electric fields near the condition of the Landau levels collapse ($\beta\rightarrow1$).
	\end{abstract}
	
	\section{Introduction}\label{intro}
The allotropes of carbon, that present a variety of crystal structures from zero- to three-dimensional, have interesting and different properties. For instance, since the isolation of graphene, the main two-dimensional allotrope of carbon, by Novoselov and Geim \cite{ng04,ng05} there has been an avalanche of theoretical and experimental research about its physical properties and technological applications~\cite{ng07,k07,cgp09,goerbig11,nbo17}. In a more general aspect, graphene has also attracted the attention to other two-dimensional Dirac materials (2D DMs)~\cite{kks09,hk10,qz11,kntsk14,jt17,ow17}. The graphene band structure is characterized by a linear behavior of the dispersion relation close to the so-called Dirac points, which are located at the Brillouin zone corners. As a consequence, its electrons are described by an effective massless Dirac equation with the Fermi velocity $v_{\rm F}\sim c/300$ playing the role of the light speed $c$. 
In the case in which an external magnetic field is applied to a graphene sample on $xy$-plane, the system that arises is described by the Dirac-Weyl (DW) equation and has been employed not only in experimental researches 
but also in several theoretical works in which different magnetic field profiles are considered (for instance, see Refs.~\cite{knn09,mf14,jk14,j15,cf19,em17}). Nevertheless, when an electric field interacts with the above system the problem nature changes, so that it is necessary to implement either a numerical method or a process that involves rotations in order to solve the equations that appear~\cite{em17,hrp10,pmp06,pc07,hr14}. For instance, for position-dependent electrostatic potentials $U(x)$, a relativistic approach is often used to become the problem into an analogous one of special relativity with a massless particle moving with an effective velocity $v_{\rm F}$ \cite{lsb07,tpl10,sgt15}. 
In general, the combined effect of magnetic and electric fields in graphene results an important research topic because it has given rise to new phenomena, such as the collapse of Landau levels \cite{pc07,lsb07,gry11,ghosh19}.

On the other hand, the physical system of a charged particle interacting with an electric and/or magnetic field is well-known in non-relativistic classical mechanics. For instance, when an electron stay in a plane and a magnetic field is applied, it describes a circular motion in the plane, while in the interaction with both magnetic and electric fields, it follows a spiral path on the plane. Thus, one would expect that it is possible to obtain a semi-classical description for the same field configuration in the graphene case by employing the so-called coherent states, which are states proposed by E. Schr\"{o}dinger~\cite{s26} as a kind of quantum states that described the motion of a particle in a quadratic potential and that minimize the Heisenberg uncertainty relation (HUR). Such states are considered as the most classical ones and have been employed and generalized to describe other physical systems, e.g., in quantum optics, atomic, nuclear, condensed matter and particle physics (see Refs.~\cite{ks85,aag00,zdh18,cbf19}). Inspired by this approach, part of this work consists of solving the problem of the interaction of a graphene sample with both magnetic and electric fields by performing a simple algebraic procedure, without the need to implement a Lorentz boost. Likewise, the coherent states construction for this case is motivated by the previous results of Refs.~\cite{df17,dnn19,cdr19} and, ultimately, it seeks to expand the theoretical background for the description of a bit different system.


This work is presented as follows. In Sec.~\ref{ECF}, we briefly discuss the classical motion equations of charged particles under the interaction of crossed electric and magnetic fields. In Sec.~\ref{DW}, the DW equation with a field configuration similar to that of the preceding section is solved. The energy spectrum and eigenvectors are found explicitly by implementing a non-relativistic approach, in contrast with previous works. In Sec.~\ref{annihilation}, a generalized annihilation operator is defined and the coherent states are constructed as its eigenstates with complex eigenvalues. In addition, the corresponding probability and current densities are described, and the effects of the electric field applied are analyzed through the Heisenberg uncertainty relation and mean energy. In Sec.~\ref{conclusions} we present our final remarks and conclusions.

\section{Charge carriers in classical fields}\label{ECF}
Let us suppose a particle of mass $m$ and charge $q$ moving on $xy$-plane interacting with crossed external electric and magnetic fields. The problem can be solved either by the Newtonian or Hamiltonian formalism in order to obtain the motion equations. By using the latter, the Hamiltonian $H_{\rm {class}}$ that describe this system is expressed as
\begin{equation}
H_{\text{class}}=\frac{1}{2m}(\mathbf{p}-\frac{q}{c}\mathbf{A})^2+qU(x),
\end{equation}
where the potentials are taking by $\mathbf{A}=Bx\hat{j}$ and $U(x)=-\mathcal{E}x$, such that $\mathbf{B}=\nabla\times\mathbf{A}=B\hat{k}$ and $\mathbf{E}=-\nabla U(x)=\mathcal{E}\hat{i}$, respectively. Considering an electron ($q=-e$), the solutions of the corresponding motion equations are given by
\begin{subequations}
	\begin{align}
x(t)&=x_{0}+\frac{1}{\omega_{\rm B}}\left[(v_{0y}+v_{\rm d})\left(\cos(\omega_{\rm B}t)-1\right)+v_{0x}\sin(\omega_{\rm B}t)\right], \\
y(t)&=y_{0}+\frac{1}{\omega_{\rm B}}\left[(v_{0y}+v_{\rm d})\sin(\omega_{\rm B}t)+v_{0x}\left(1-\cos(\omega_{\rm B}t)\right)\right]-v_{\rm d}t,
\end{align}
\end{subequations}
where $\mathbf{r}_{0}=(x_{0},y_{0})$ is the initial position of the particle, $\mathbf{v}_{0}=(v_{0x},v_{0y})$ denotes its initial velocity, $\omega_{\rm B}=eB/mc$ is the cyclotron frequency and $v_{\rm d}=c\,\mathcal{E}/B$ is the drift velocity. 

\begin{figure}[tb]
	\centering
	\includegraphics[width=0.7\linewidth]{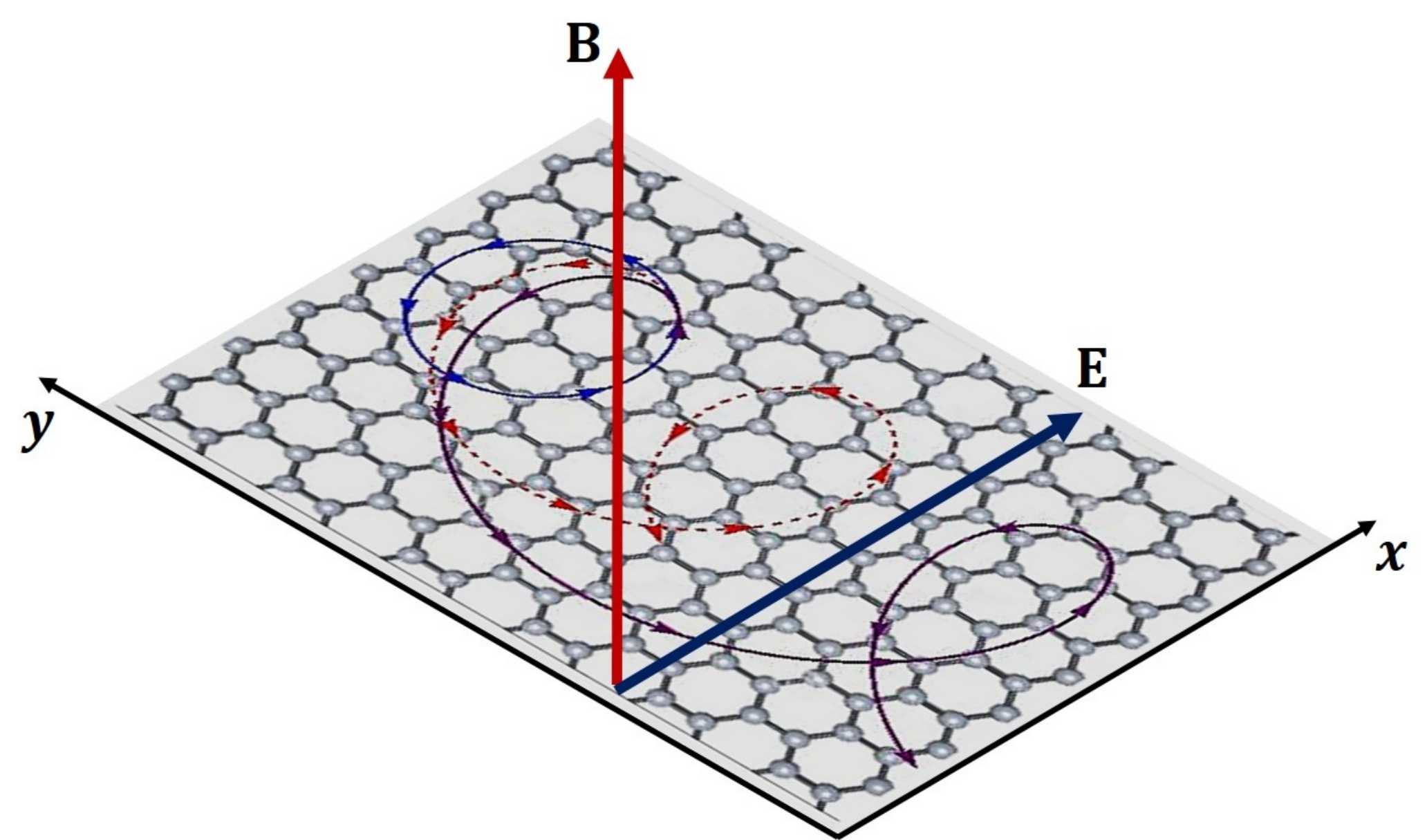}
	\caption{Graphene layer interacting with crossed electric and magnetic fields directed along the $x$- and $z$-directions, respectively. In absence of electric field $\mathbf{E}$, a classical charged particle performs a circular trajectory (blue curve), while if the strength $\mathcal{E}$ increases, the trajectory becomes into a trochoid (red and purple curves).}
	\label{fig:graphene}
\end{figure}

From the above equations, we obtain the expression
\begin{equation}\label{4}
	\left(x(t)-x_{0}+\frac{v_{0y}+v_{\rm d}}{\omega_{\rm B}}\right)^2+\left(y(t)-y_{0}+v_{\rm d}t-\frac{v_{0x}}{\omega_{\rm B}}\right)^2=\frac{1}{\omega_{\rm B}^2}\left[\left(v_{0y}+v_{\rm d}\right)^2+v_{0x}^2\right],
\end{equation} 
that corresponds to the equation of a circumference centered in the point
\begin{equation}
(h,k)=\left(x_{0}-\frac{v_{0y}+v_{\rm d}}{\omega_{\rm B}},y_{0}-v_{\rm d}t+\frac{v_{0x}}{\omega_{\rm B}}\right),
\end{equation}
and radius $R=\omega_{\rm B}^{-1}\sqrt{\left(v_{0y}+v_{\rm d}\right)^2+v_{0x}^2}$. Here, the circle center moves along the $y$-axis with speed $v_{\rm d}$. When the electric field $\mathbf{E}$ is null, we have that $v_{\rm d}=0$ and Eq.~(\ref{4}) corresponds to circle equation with center in $\left(x_{0}-v_{0y}/\omega_{\rm B},y_{0}+v_{0x}/\omega_{\rm B}\right)$ and radius $R=\vert\mathbf{v}_{0}\vert/\omega_{\rm B}$. In presence of an electric field $\mathbf{E}$, instead of a circular motion, the trajectories describe a cycloidal motion along the $y$-axis. If the strength $\mathcal{E}$ increases, the trajectories open further towards the negative $x$-axis (see Fig.~\ref{fig:graphene}).

On the other hand, the dynamics of relativistic particles is studied through appropriate Lorentz transformations that allow to find the motion equations in a coordinate frame $S'$ moving with a velocity $\mathbf{u}$ with respect to the original frame $S$. For the case with $\mathcal{E}<B$, the velocity $\mathbf{u}$ is chosen perpendicular to the vectors $\mathbf{E}$ and $\mathbf{B}$, so that the only field acting in the frame $S'$ is $\mathbf{B}'$, and $\mathbf{u}$ has a physical meaning as the drift velocity if $\vert\mathbf{u}\vert$ is less than $c$. For $\mathcal{E}>B$, it is necessary to perform a different Lorentz transformation from the frame $S$ to a system $S''$ in which the only field acting is now $\mathbf{E}'$ that causes a hyperbolic motion~\cite{jackson99}.

\section{Dirac-Weyl equation with external fields}\label{DW}
Now, let us consider a graphene layer laying on $xy$-plane interacting with external electric and magnetic fields, that are respectively parallel and orthogonal to the layer surface, as illustrated in Fig.~\ref{fig:graphene}. The Hamiltonian that describes this problem is given by
\begin{equation}\label{5}
H\bar{\Psi}(x.y)=\left[v_{\rm F}\bm{\sigma}\cdot\left(\mathbf{p}+\frac{e}{c}\mathbf{A}\right)-eU(x)\right]\bar{\Psi}(x,y)=E\bar{\Psi}(x,y),
\end{equation}
where $\bar{\Psi}(x,y)=\exp\left(iky\right)\Psi(x)$, the electrostatic and magnetic potentials are given again by $U(x)=-\mathcal{E}x$ and $\mathbf{A}(x)=Bx\hat{j}$, respectively, and $\bm{\sigma}=(\sigma_{x},\sigma_{y})$ are the Pauli matrices.

Thus, the above equation becomes:
\begin{align}\label{91}
&\left[\left(\frac{E-e\mathcal{E}x}{\hbar v_{\rm F}}\right)\mathbb{I}_{2}+i\partial_x\sigma_{x}-\frac{1}{l_{\rm B}^2}\left(x+l_{\rm B}^2k\right)\sigma_{y}\right]\Psi(x)=0,
\end{align}
where $\mathbb{I}_{2}$ denotes the $2\times2$ unity matrix and the magnetic length $l_{\rm B}$ is given by $l_{\rm B}^2=2/\omega_{\rm B}=c\hbar/eB$.
By introducing the parameter $\beta$ and the dimensionless quantity $\xi$,
\begin{equation}
\beta=\frac{e\mathcal{E}}{\hbar v_{\rm F}}l_{\rm B}^2=\frac{c\mathcal{E}}{v_{\rm F}B}=\frac{v_{\rm d}}{v_{\rm F}}, \quad \xi=\frac{1}{l_{\rm B}}\left(x+l_{\rm B}^2k\right),
\end{equation}
Eq.~(\ref{91}) can be rewritten as
\begin{equation}\label{8}
\left[\left(\frac{E}{\hbar v_{\rm F}}+k\beta-\frac{\beta \xi}{l_{\rm B}}\right)\mathbb{I}_{2}+\frac{i}{l_{\rm B}}\frac{d}{d\xi}\sigma_{x}-\frac{\xi}{l_{\rm B}}\sigma_{y}\right]\Psi(\xi)=0,
\end{equation}
which is solved in detail on Appendix~\ref{eigenstates}.

\begin{figure}[tb]
	\centering
	\includegraphics[width=0.55\linewidth]{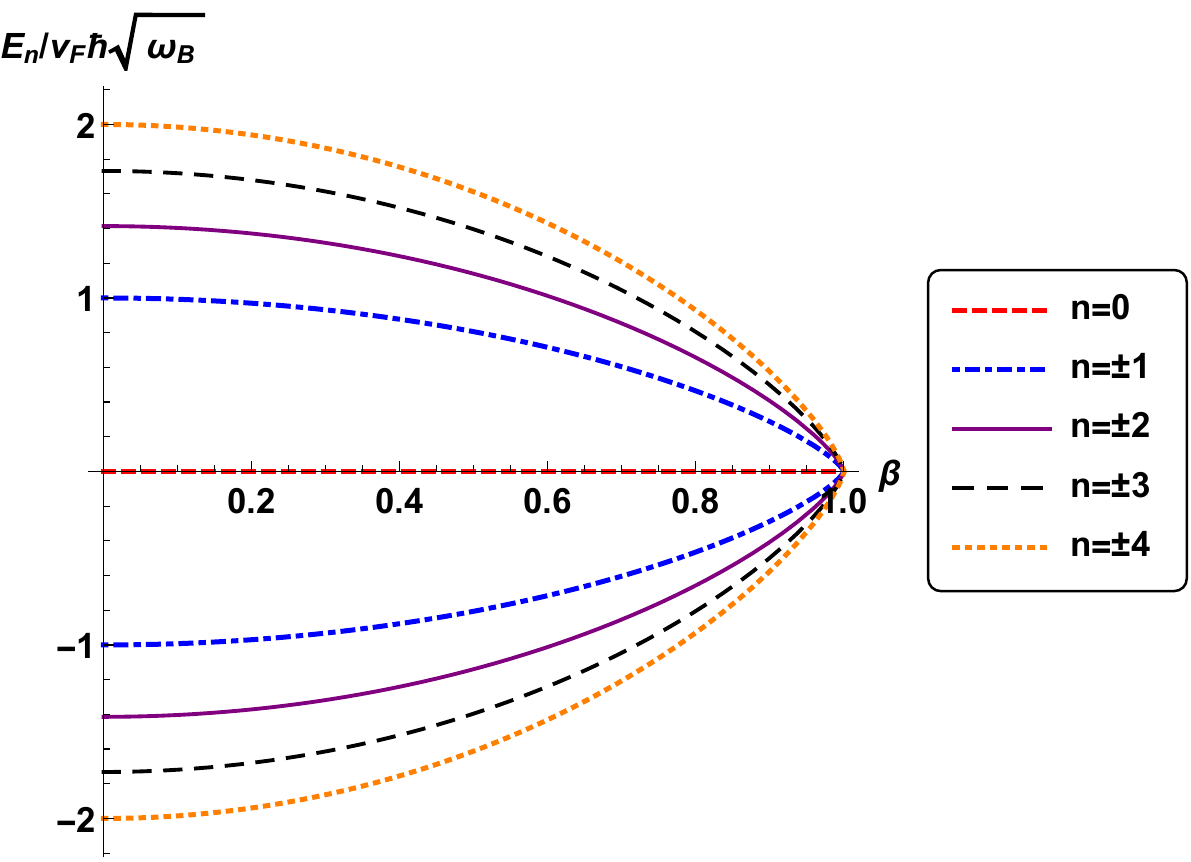}
	\caption{Plot of Landau levels (LLs) as a function of $\beta$ and $k=0$. For $\beta=0$, the energy eigenstates lie on LLs, but as $\beta\rightarrow1$, Landau level spectrum collapses.}
	\label{fig:LL}
\end{figure}

The energy spectrum $E_n$ is given by \cite{goerbig11,lsb07,pc07,sgt15} (see Fig.~\ref{fig:LL}):
\begin{equation}\label{energy}
E_{(1,n-1)}=E_{(2,n)}=\text{sgn}(n)\frac{\hbar v_{\rm F} (1-\beta^2)^{3/4}}{l_{\rm B}}\sqrt{2n}-\hbar v_{\rm F} k\beta,
\end{equation}
where sgn$(0)=1$, while the eigenspinors $\bar{\Psi}_{n}(x,y)$ can be expressed as:
\begin{align}\label{20}
\nonumber\bar{\Psi}_{n}(x,y)\equiv\bar{\Psi}_{n}(\zeta_{n},y)&=\frac{\exp\left(iky\right)}{\sqrt{2^{(1-\delta_{0n})}}}\left[(1-\delta_{0n})\psi_{n-1}(\zeta_{n})\chi_{\lambda_1}+\eta\,\psi_{n}(\zeta_{n})\chi_{\lambda_2}\right]\\
\nonumber&=\frac{\exp\left(iky\right)}{\sqrt{2^{(1-\delta_{0n})}}}\sqrt{\frac{1}{2}}\left(\begin{array}{c c}
\sqrt{C_+} & i\sqrt{C_-} \\
-i\sqrt{C_-} &  \sqrt{C_+} \\
\end{array}\right)\left(\begin{array}{c}
(1-\delta_{0n})\psi_{n-1}(\zeta_{n}) \\
i\eta\,\psi_{n}(\zeta_{n})
\end{array}\right) \\
&=\mathbb{M}\bar{\Phi}_{n}(x,y), \quad n=0,1,2,\dots,
\end{align}
Here, $\delta_{mn}$ denotes the Kronecker delta, $\eta=+$ for the $K$ valley while $\eta=-$ for the $K'$ valley,
\begin{subequations}\label{11}
	\begin{align}
\mathbb{M}&=\sqrt{\frac{1}{2}}\left(\begin{array}{c c}
\sqrt{C_+} & i\sqrt{C_-}\\
-i\sqrt{C_-} & \sqrt{C_+}
\end{array}\right)=\sqrt{\frac{1}{2}}\left(\sqrt{C_+}\mathbb{I}_{2}-\sqrt{C_-}\sigma_{y}\right), \\ 
\bar{\Phi}_{n}(x,y)&=\frac{\exp\left(iky\right)}{\sqrt{2^{(1-\delta_{0n})}}}\left(\begin{array}{c}
(1-\delta_{0n})\psi_{n-1}(\zeta_{n}) \\
i\eta\,\psi_{n}(\zeta_{n})
\end{array}\right),
\end{align}
\end{subequations}
with $C_{\pm}=1\pm(1-\beta^2)^{1/2}$ and the wave functions are given by
\begin{equation}\label{106}
\psi_{n}(x)\equiv\psi_{n}(\zeta_{n})=\frac{(1-\beta^2)^{1/8}}{\sqrt{n!}}\left(\frac{\omega_{\rm B}}{2\pi}\right)^{1/4}D_n(\sqrt{2}\,\zeta_{n}),
\end{equation}
where $D_n(z)=U\left(-n-\frac12;z\right)$ designates the parabolic cylinder function and
\begin{equation}
\zeta_{n}=\xi(1-\beta^2)^{1/4}-\frac{\beta\epsilon_0l_{\rm B}}{(1-\beta^2)^{3/4}}=\frac{(1-\beta^2)^{1/4}}{l_{\rm B}}\left[x+l_{\rm B}^2k+\text{sgn}(n)\frac{\beta l_{\rm B}\sqrt{2n}}{(1-\beta^2)^{1/4}}\right].
\end{equation}

In Eq.~(\ref{energy}), positive (negative) energies correspond to Dirac fermions in the conduction (valence) band. For $\beta=0$ ({\it i.e.,} vanishing electric field), the spinors $\bar{\Phi}_{n}$ in~(\ref{11}) reduce to the solutions of standard Landau-like problem that was considered in Refs.~\cite{ng04,ng05,knn09,mf14,df17}, for instance. A brief comment about matrix $\mathbb{M}$ can be found in Appendix~\ref{matrixM}.

\begin{figure}[!ht]
	\centering
	\includegraphics[width=0.9\textwidth]{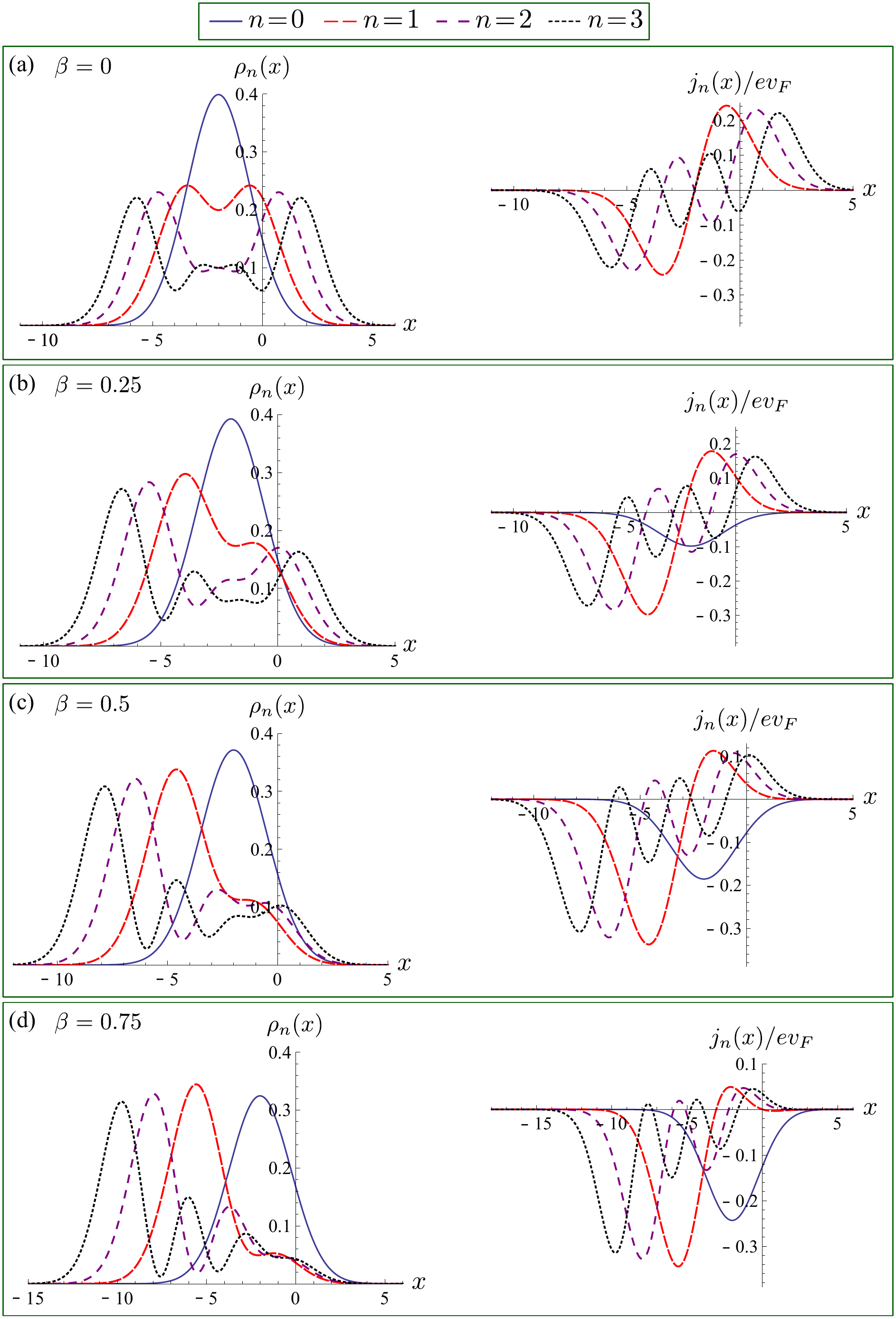}
	\caption{\label{fig:rhoN}Upper panels show the probability density $\rho_n(x)$ (left-hand) and the $y$-current density $j_{n}(x)/e\,v_{\rm F}$ (right-hand) for different eigenstates $\bar{\Phi}_{n}$ and values of $\beta$. In all these cases $B=1/2$, $k=\omega_{\rm B}=1$.}
\end{figure}

Figure~\ref{fig:rhoN} shows the behavior of the probability density $\rho_{n}(x)$ and $y$-current density $j_{y}(x)$ which were built with the eigenstates in~(\ref{20}) (see \ref{densities}). In particular, for $\beta=0$, $\rho_{n}(x)$ is an even function while $j_{y}(x)$ is an odd function respect to reflection around a point $x_{0}$ such that $\xi\vert_{x=x_{0}}=0$. However, as $\beta\rightarrow1$ the parity breaks and the maximum values of both functions move in the negative $x$-direction.  
Furthermore, according to \cite{knn09} the sign of $j_{y}(x)$ indicates the direction along the $y$-axis in which the electron motion takes place, so the current density behavior suggests there is a flux of probability in the negative $y$-direction when an electric field is applied to the sample. It is worth to mention that Figs.~\ref{fig:rhoN}{\color{blue}a} and \ref{fig:rhoN}{\color{blue}b} are according with the results on \cite{knn09}.

Finally, according to \cite{ziman72,huang18} the average velocity in the $y$-direction is given by
\begin{equation}
\langle v_{y}\rangle=\frac1\hbar\frac{\partial E_{n}}{\partial k}=-v_{\rm F}\beta=-v_{\rm d}=\left[\frac{\mathbf{E}\times\mathbf{B}}{B^2}\right]_{y},
\end{equation}
that means the Dirac fermions move with an average velocity $v_{\rm d}$ in the negative $y$-direction.

\section{Annihilation operator}\label{annihilation}
In order to build the coherent states associated with the problem, it is necessary to 
perform a transformation that allows us to work with an adequate set of eigenstates for which can be defined an annihilation operator, namely $\Theta^-$. Thus, considering the inverse matrix $\mathbb{M}^{-1}$, we obtain:
\begin{equation}
\bar{\Phi}_{n}(x,y)=\mathbb{M}^{-1}\bar{\Psi}_{n}(x,y)=\frac{\exp\left(iky\right)}{\sqrt{2^{(1-\delta_{0n})}}}\left(\begin{array}{c}
(1-\delta_{0n})\psi_{n-1}(\zeta_{n}) \\
i\eta\,\psi_{n}(\zeta_{n})
\end{array}\right).
\end{equation}\label{31}

In this representation, we can define the following differential operators
	\begin{equation}
	\theta_{n}^{\pm}=\frac{1}{\sqrt{2}}\left(\mp\frac{d}{d\zeta_{n}}+\zeta_{n}\right), \quad \theta_{n}^{+}=(\theta_{n}^{-})^{\dagger},
	\end{equation}
	as well as the (unitary) shift operators $\mathcal{T}^{\pm}$~\cite{golinshi19}, 
whose explicit action onto the eigenfunctions $\psi_{n}(\zeta_{n})$ is
	\begin{equation}
	\mathcal{Q}^{-}\psi_{n}(\zeta_{n})=\mathcal{T}^{-}\theta_{n}^{-}\psi_{n}(\zeta_{n})=\sqrt{n}\,\psi_{n-1}(\zeta_{n-1}), \quad \mathcal{Q}^{+}\psi_{n}(\zeta_{n})=\theta_{n}^{+}\mathcal{T}^{+}\psi_{n}(\zeta_{n})=\sqrt{n+1}\,\psi_{n+1}(\zeta_{n+1}).
	\end{equation}
	This means that the operators $\theta_{n}^{\pm}$ change in a unity the energy level of the eigenstates $\psi_{n}$, while the operators $\mathcal{T}^{\pm}$ shift in a unity the index $n$ of the spacial coordinate $\zeta_{n}$ in the wave function $\psi_{n}(\zeta_{n})$. It is straightforward to verify that $[\mathcal{Q}^{-},\mathcal{Q}^{+}]=1$.


Now, we define the following operators in terms of $\theta_{n}^{\pm}$ \cite{df17,cdr19}:
\begin{equation}\label{ladder}
\Theta_{n}^{-}=\left(\begin{array}{c c}
\cos(\delta)\frac{\sqrt{N+2}}{\sqrt{N+1}}\theta_{n}^- & \eta\sin(\delta)\frac{1}{\sqrt{N+1}}(\theta_{n}^-)^2 \\
-\eta\sin(\delta)\sqrt{N+1} & \cos(\delta)\theta_{n}^-
\end{array}\right), \quad \Theta_{n}^+=(\Theta_{n}^-)^\dagger.
\end{equation}
Here, $N=\theta_{n}^+\theta_{n}^-$ is the number operator and $\delta\in[0,2\pi]$ is a parameter that allows to work with either diagonal or non-diagonal matrix operators. This annihilation operator coincides with a given one in~\cite{df17,cdr19}.
After that, we also define the matrix operators
	\begin{equation}
	\Theta^{-}=\mathcal{T}^{-}\sum_{n=0}\Theta_{n}^{-}\mathcal{P}(n), \quad \Theta^+=(\Theta^-)^\dagger,
	\end{equation}
	where $\mathcal{P}(k)$ is a 1-dimensional projection such that~\cite{isozaki12}
	\begin{equation}
	\Theta^{-}\bar{\Phi}_{k}=\mathcal{T}^{-}\sum_{n=0}\Theta_{n}^{-}(\mathcal{P}(n)\bar{\Phi}_{k})=\mathcal{T}^{-}\sum_{n=0}\delta_{kn}\Theta_{n}^{-}\bar{\Phi}_{k}=\mathcal{T}^{-}\Theta_{k}^{-}\bar{\Phi}_{k}.
	\end{equation}
	We discuss the algebraic relations of these matrix operators in Appendix~\ref{ladderop}.


\subsection{Coherent states as eigenstates of $\Theta^{-}$}
The action of the annihilation operator $\Theta^-$ on the eigenstates $\bar{\Phi}_{n}(x,y)$ turns out to be
	\begin{equation}\label{35}
	\Theta^-\bar{\Phi}_{n}(x,y)\equiv\Theta^-\bar{\Phi}_{n}(\zeta_{n},y)=\frac{\exp\left(i\,\delta\right)}{\sqrt{2^{\delta_{1n}}}}\sqrt{n}\bar{\Phi}_{n-1}(\zeta_{n-1},y), \quad n=0,1,2,\dots.
	\end{equation}

\begin{figure}[!ht]
	\centering
	\includegraphics[width=0.74\textwidth]{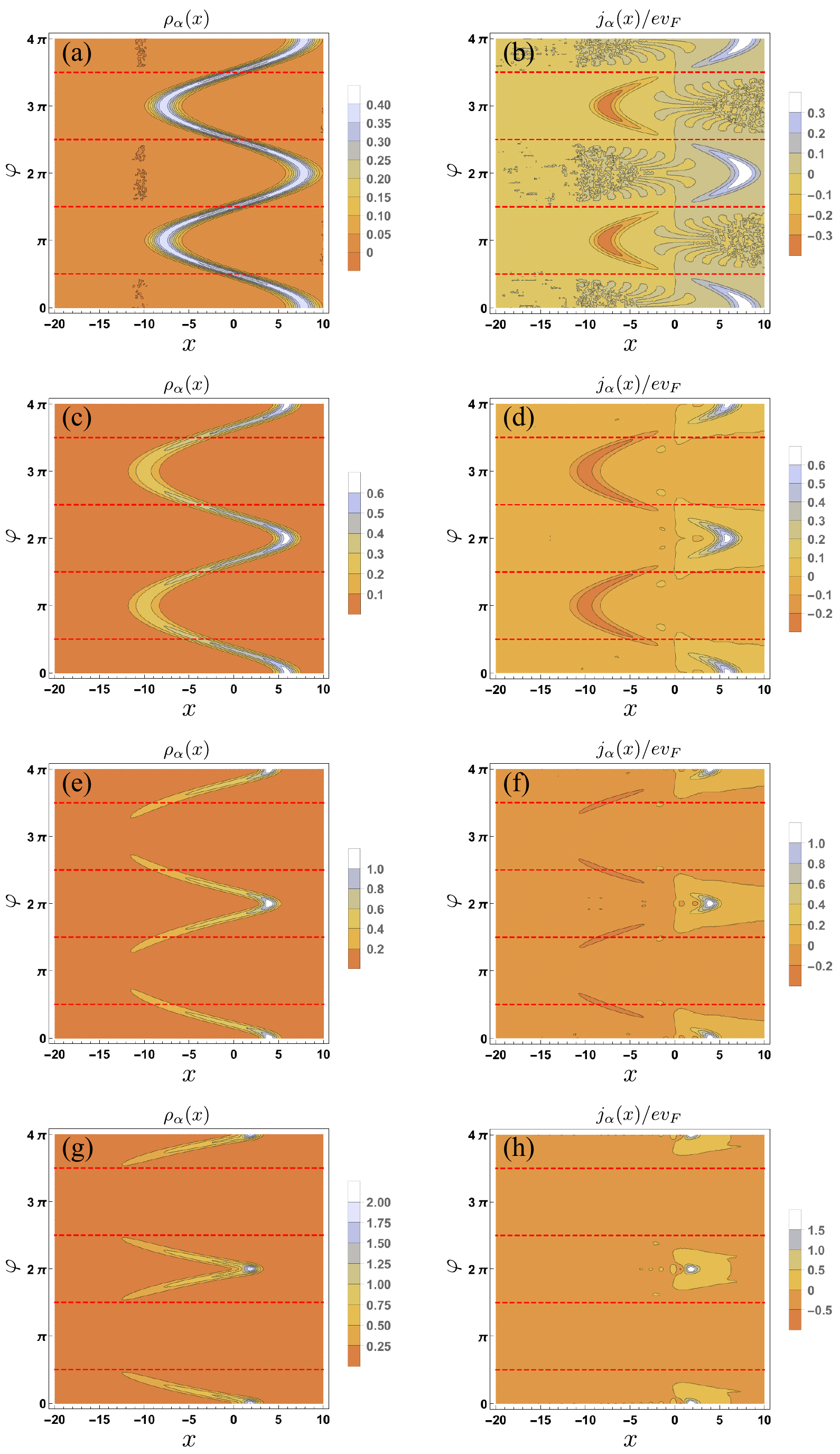}
	\caption{\label{fig:rhojy}Upper panels show the probability density $\rho_{\alpha}(x)=\vert\Psi_{\alpha}(x,y)\vert^2$ (left-hand) and the $y$-current density $j_{\alpha}(x)/e\,v_{\rm F}$ (right-hand) in~(\ref{density}) for the coherent states $\Psi_{\alpha}(x,y)$ as functions of the phase $\varphi$ for different values of $\beta$: $\beta=0$ (a, b), $\beta=0.25$ (c, d), $\beta=0.5$ (e, f) and $\beta=0.75$ (g, h). In all these cases $\vert\alpha\vert=4$, $B=1/2$, $\omega_{\rm B}=1$ and $k=\delta=0$. Dashed red lines correspond to $\varphi=(2n+1)\pi/2$, $n=0,1,2,\dots$.}
\end{figure}

By solving the eigenvalue equation
\begin{equation}
\Theta^-\Phi_{\alpha}(x,y)=\alpha\Phi_{\alpha}(x,y), \quad \alpha\in\mathbb{C},
\end{equation}
where
\begin{equation}
\Phi_{\alpha}(x,y)=\sum_{n=0}^{\infty}a_{n}\bar{\Phi}_{n}(x,y),
\end{equation}
and using~(\ref{35}), we find the explicit expression for the corresponding coherent states:
\begin{align}\label{40}
\Phi_{\alpha}(x,y)=\frac{1}{\sqrt{2\exp\left(\vert \tilde{\alpha}\vert^2\right)-1}}\left[\bar{\Phi}_{0}(x,y)+\sum_{n=1}^{\infty}\frac{\sqrt{2}\tilde{\alpha}^n}{\sqrt{n!}}\bar{\Phi}_{n}(x,y)\right],
\end{align}
with $\tilde{\alpha}\equiv\alpha\exp(-i\,\delta)$, and $\alpha=\vert\alpha\vert\exp\left(i\varphi\right)$. This means that to work with either a diagonal or non-diagonal annihilation operator $\Theta^-$ results in the introduction of a phase factor that affects the eigenvalue $\alpha$. Actually, the phase of $\alpha$ already corresponds to the classical phase angle \cite{cn65,nms96}, {\it i.e.}, $\alpha$ is a periodic amount that contains information about the cyclic change of the mean value of both the position and the momentum, as will be seen in forthcoming sections. 

\subsubsection{Probability and current densities}
The corresponding probability and $y$-current densities for the coherent states $\Psi_{\alpha}=\mathbb{M}\Phi_{\alpha}$ read as (see Fig.~\ref{fig:rhojy}):


\footnotesize
\begin{subequations}
\begin{align}\label{density}
\nonumber&\vert\Psi_{\alpha}(x,y)\vert^2=\left[2\exp\left(\vert\tilde{\alpha}\vert^2\right)-1-2\beta\Re\left(\tilde{\alpha}\right)\sum_{n=0}^{\infty}\frac{\vert\tilde{\alpha}\vert^{2n}}{n!\sqrt{n+1}}\right]^{-1}\Bigg[\psi_{0}^2(\zeta_{0})+\left\vert\sum\limits_{n=1}^{\infty}\frac{\tilde{\alpha}^n}{\sqrt{n!}}\psi_{n}(\zeta_{n})\right\vert^2+\left\vert\sum\limits_{n=1}^{\infty}\frac{\tilde{\alpha}^n}{\sqrt{n!}}\psi_{n-1}(\zeta_{n})\right\vert^2\nonumber\\
&\quad+2\Re\left(\sum\limits_{n=1}^{\infty}\frac{\tilde{\alpha}^n}{\sqrt{n!}}\psi_{n}(\zeta_{n})\psi_{0}(\zeta_{0})\right)-2\beta\Re\left(\sum\limits_{n=1}^{\infty}\frac{\tilde{\alpha}^n}{\sqrt{n!}}\psi_{n-1}(\zeta_{n})\psi_{0}(\zeta_{0})+\sum_{m,n=1}^{\infty}\frac{\tilde{\alpha}^{\ast m}\tilde{\alpha}^n}{\sqrt{m!\,n!}}\psi_{m-1}(\zeta_{n})\psi_{n}(\zeta_{n})\right)\Bigg], \\
\nonumber&\frac{j_{\alpha}(x)}{ev_{\rm F}}=\left[2\exp\left(\vert\tilde{\alpha}\vert^2\right)-1-2\beta\Re\left(\tilde{\alpha}\right)\sum_{n=0}^{\infty}\frac{\vert\tilde{\alpha}\vert^{2n}}{n!\sqrt{n+1}}\right]^{-1}\Bigg\{2\Re\Bigg(\sum\limits_{n=1}^{\infty}\frac{\tilde{\alpha}^n}{\sqrt{n!}}\psi_{n-1}(\zeta_{n})\psi_{0}(\zeta_{0})+\sum_{m,n=1}^{\infty}\frac{\tilde{\alpha}^{\ast m}\tilde{\alpha}^n}{\sqrt{m!\,n!}}\psi_{m-1}(\zeta_{n})\\
&\quad\times\psi_{n}(\zeta_{n})\Bigg)-\beta\left[\psi_{0}^2(\zeta_{0})+\left\vert\sum\limits_{n=1}^{\infty}\frac{\tilde{\alpha}^n}{\sqrt{n!}}\psi_{n}(\zeta_{n})\right\vert^2+\left\vert\sum\limits_{n=1}^{\infty}\frac{\tilde{\alpha}^n}{\sqrt{n!}}\psi_{n-1}(\zeta_{n})\right\vert^2+2\Re\left(\sum\limits_{n=1}^{\infty}\frac{\tilde{\alpha}^n}{\sqrt{n!}}\psi_{n}(\zeta_{n})\psi_{0}(\zeta_{0})\right)\right]\Bigg\}.
\end{align}
\end{subequations}
\normalsize



For $\beta=0$ ($\mathcal{E}=0$), we recover the results discussed in~\cite{df17}. 
For $\beta\neq0$, we can see that the shape of both probability and $y$-current densities changes as $\beta$ increases. In a semi-classical interpretation, the eigenvalue $\alpha$ translates as an initial condition with which we can indicate the initial position of an electron along the $x$-axis for a given time $t_{0}$. On the other hand, when the amount $\beta\rightarrow1$, the probability density shows a maximum value in a small region in the $x$-axis, {\it i.e.}, such behavior suggests that the presence of an electric field would tend to reduce the velocity of the electrons in particular regions along the $x$ axis, so that the probability of finding them there will increase. Meanwhile, the sign of the $y$-current density indicates the direction in which the electron movement takes place along the $y$-direction \cite{knn09}. For $\beta=0$, such a function has both positive and negative values, so that the electrons move as many times to the positive as the negative direction along the $y$-axis. As $\beta\rightarrow1$, small regions in which the current density becomes most positive appear, indicating that when the electron speed reduces, it moves in the positive $y$-direction.

\begin{figure}[!ht]
	\centering
	\includegraphics[width=0.95\textwidth]{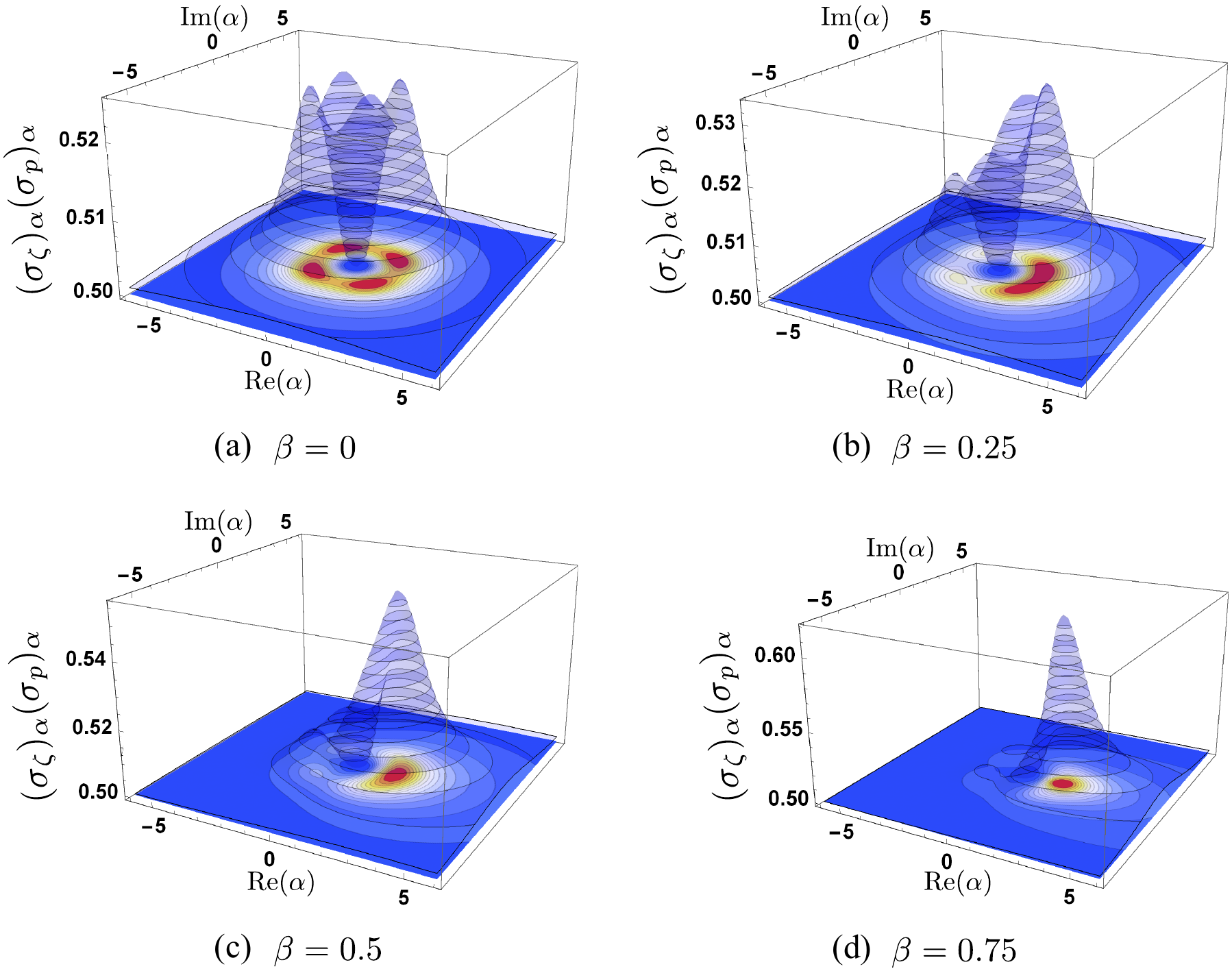}
	\caption{\label{fig:HUR}HUR for the coherent states $\Psi_{\alpha}(x,y)$ as function of the eigenvalue $\alpha$ with $\delta=0$ for some values of $\beta$.}
\end{figure}

\begin{figure}[!ht]
	\centering
	\includegraphics[width=0.9\textwidth]{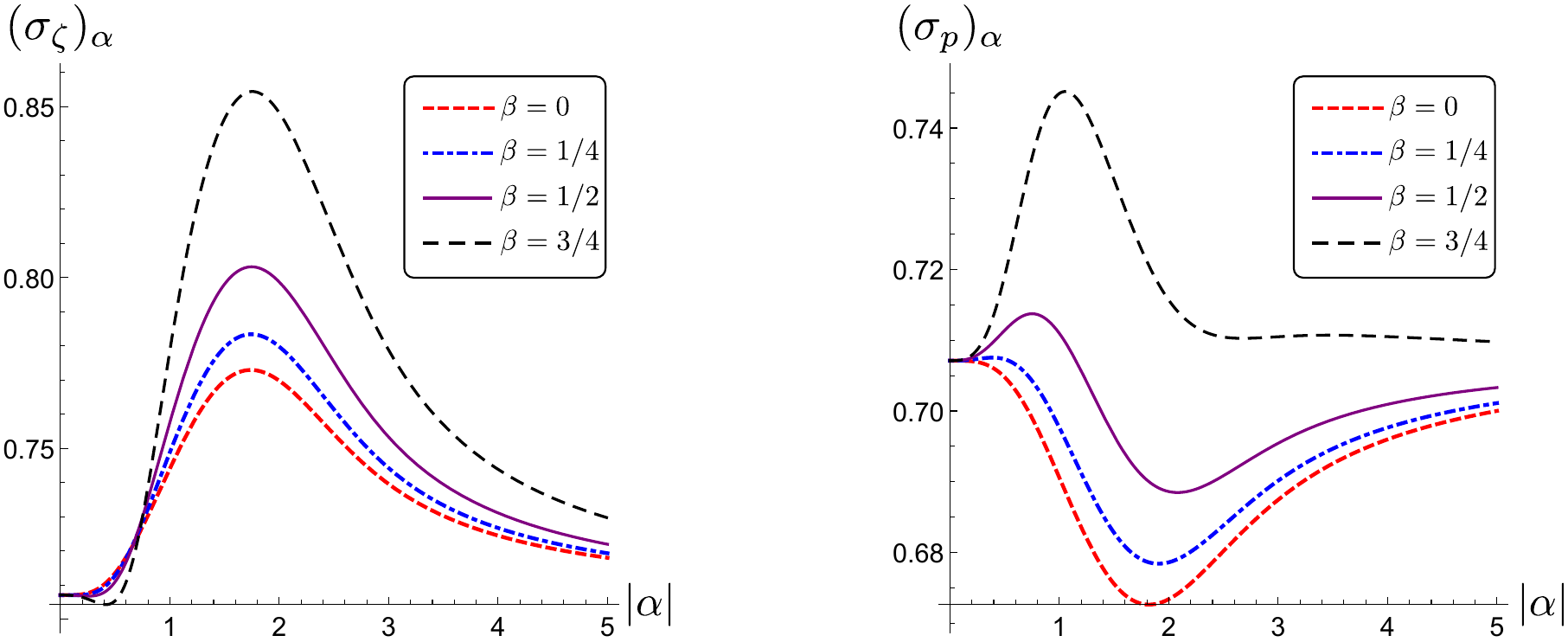}
	\caption{\label{fig:HURCL}Comparison between the position and momentum variances as function of $\vert\alpha\vert$ for different values of $\beta$. In general, $\sigma_{p}\leq\sigma_{\zeta}$ and as $\beta\rightarrow1$, $\sigma_{\zeta}$ increases.}
\end{figure}

\subsubsection{Heisenberg uncertainty relation}
In order to compute the Heisenberg uncertainty relation, we define the following matrix quadrature $\mathbb{S}_q$ and its square as
\begin{equation}
	\mathbb{S}_q=\sum_{n=0}(s_q\otimes\mathbb{I})\mathcal{P}(n), \quad \mathbb{S}_q^2=\sum_{n=0}(s_q^2\otimes\mathbb{I})\mathcal{P}(n),
	\end{equation}
	where
	\begin{subequations}
		\begin{eqnarray}
		&&s_q=\frac{1}{\sqrt{2}i^q}\left(\mathcal{Q}^-+(-1)^q\mathcal{Q}^+\right), \\
		&&s_q^2=\frac{1}{2}\left[2N+1+(-1)^q((\mathcal{Q}^-)^2+(\mathcal{Q}^+)^2)\right],
		\end{eqnarray}
\end{subequations}
and $q=0,1$. The variance of the operator $\mathbb{S}_q$ is calculated as follows:
\begin{equation}
\sigma_{\mathbb\mathbb{S}_q}=\sqrt{\langle\mathbb{S}_q^2\rangle-\langle\mathbb{S}_q\rangle^2}.
\end{equation}
The explicit expressions of $\langle\mathbb{S}_q\rangle$ and $\langle\mathbb{S}_q^2\rangle$ in the coherent states basis are given in Appendix~\ref{heisenberg}. For $q=0$ ($q=1$), we have that $\sigma_{\mathbb{S}_0}\equiv\sigma_{\zeta}$ ($\sigma_{\mathbb{S}_1}\equiv\sigma_{p}$), {\it i.e.}, the variance of the position-like $\zeta$ (momentum-like $p$) operator that fulfills:
	\begin{equation}
	\sigma_{\zeta}\sigma_{p}=\sigma_{\mathbb{S}_0}\sigma_{\mathbb{S}_1}\geq\frac12.
	\end{equation}
For the standard coherent states (SCS) of the harmonic oscillator, the equality is verified.

\begin{figure}[!ht]
	\centering
	\includegraphics[width=\textwidth]{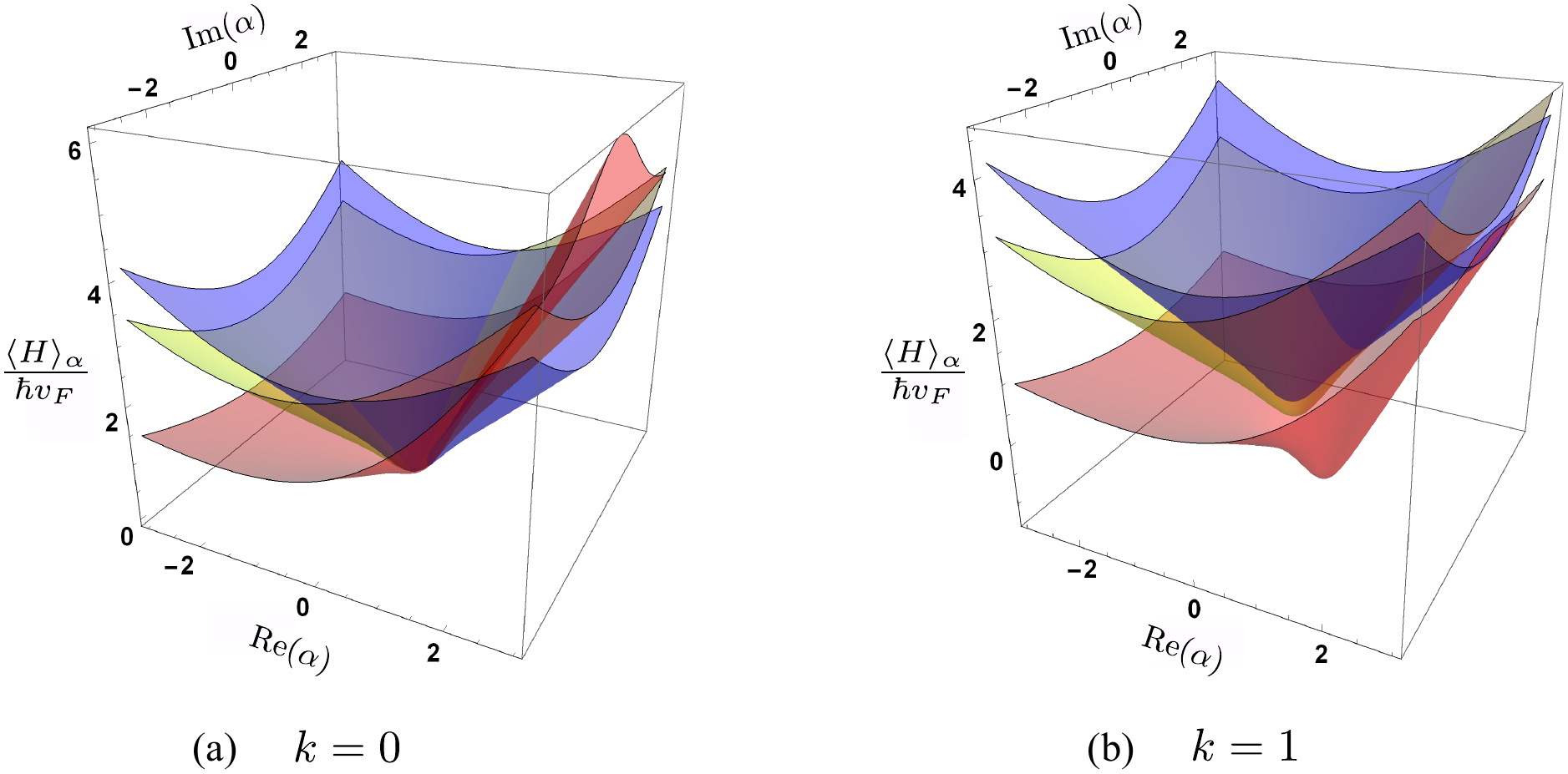}
	\caption{\label{fig:Hbeta}Mean energy $\langle H\rangle_{\alpha}/\hbar v_{\rm F}$ as function of $\alpha$ with $\delta=0$ for the coherent states $\Psi_{\alpha}(x,y)$ with different values of $\beta$: $\beta=0$ (blue), $\beta=0.25$ (yellow) and $\beta=0.75$ (red). In all these cases, we take $B=1/2$ and $\omega_{\rm B}=1$.}
\end{figure}

Figure~\ref{fig:HUR} shows that for coherent states $\Psi_{\alpha}=\mathbb{M}\Phi_{\alpha}$, the function $\sigma_{\zeta}\sigma_{p}$ is greater than $1/2$ for some values of $\alpha$ and $\varphi$. In general, the behavior of $\sigma_{\zeta}\sigma_{p}$ for small values of $\vert\alpha\vert$ depends on the individual variances of the quadratures $\mathbb{S}_{0}$ (position-like operator) and $\mathbb{S}_{1}$ (momentum-like operator) on the coherent state basis considered, as shown in figure 5 of Ref.~\cite{cdr19}. Additionally, for $\varphi$ close to zero and growing $\beta$, the variance $\sigma_{p}$ remains less than $\sigma_{\zeta}$ and when $\beta\sim1$, $\sigma_{\zeta}\sigma_{p}$ reaches its maximum value in the vicinity of $\varphi=0$ and for small values of $\vert\alpha\vert$ (see Fig.~\ref{fig:HURCL}).

\subsubsection{Mean energy value}
Recalling that for any linear combination of eigenstates $\Psi_{n}$ of a Hamiltonian $H'$ with eigenvalues $E_{n}$ we have that
\begin{equation}
\Psi=\sum_{n}a_{n}\Psi_{n}\quad \Longrightarrow\quad \langle H'\rangle=\sum_{n}\vert a_{n}\vert^2E_{n},
\end{equation}
the mean energy $\langle H\rangle_{\alpha}$ for the coherent states $\Psi_{\alpha}(x,y)$ can be expressed as follows:
\begin{align}\label{39}
\nonumber\frac{\langle H\rangle_{\alpha}}{\hbar v_{\rm F}}&=\left[2\exp\left(\vert\tilde{\alpha}\vert^2\right)-1-2\beta\Re\left(\tilde{\alpha}\right)\sum_{n=0}^{\infty}\frac{\vert\tilde{\alpha}\vert^{2n}}{n!\sqrt{n+1}}\right]^{-1}\Bigg[k\beta\left(1-2\exp\left(\vert\tilde{\alpha}\vert^2\right)\right)\\
&\quad+\frac{2(1-\beta^2)^{3/4}}{l_{\rm B}}\sum_{n=1}^{\infty}\frac{\vert\tilde{\alpha}\vert^{2n}}{n!}\sqrt{2n}\,\Bigg].
\end{align}

As Fig.~\ref{fig:Hbeta} shows, $\langle H\rangle_{\alpha}$ is a continuous function of the eigenvalue $\alpha$, that allows us to assure the semi-classical nature of our results. Furthermore, the mean energy value exhibits a conic shape around the eigenvalue $\alpha=0$, whose inclination increases as $\beta$ also does. Actually, this behavior suggests that the presence of an external electric field works as a parameter of tilt for the mean energy. Such behavior is common in 2D DMs that possess tilted (anisotropic) Dirac cones \cite{ow17,sgt15,stg14}.

\begin{figure}[!ht]
	\centering
	\includegraphics[width=0.5\textwidth]{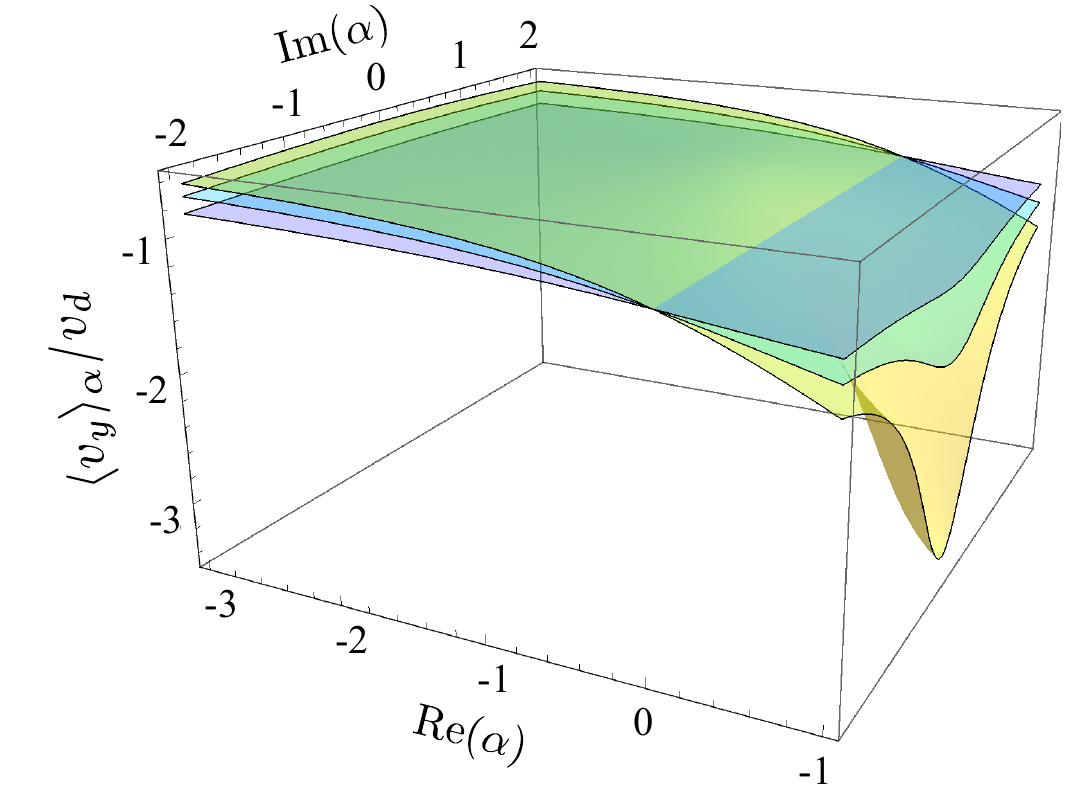}
	\caption{\label{fig:vy}Average velocity in $y$-direction $\langle v_{y}\rangle_{\alpha}/v_{\rm d}$ as function of $\alpha$ with $\delta=0$ for the coherent states $\Psi_{\alpha}(x,y)$ with different values of $\beta$: $\beta=0.25$ (blue), $\beta=0.5$ (cyan) and $\beta=0.75$ (yellow). In all these cases, we take $B=1/2$ and $\omega_{\rm B}=1$.}
\end{figure}

\subsection{Discussion}

In comparison with~\cite{df17}, we also obtain an oscillatory-like behavior for the probability and current densities around a middle point $x_0$, whose location on $x$-axis depends on the strength of $\mathbf{E}$ and $\mathbf{B}$. For the case with the electric field turned off, the function $\rho_{\alpha}(x)$ has only contributions of the spinor components in each sublattice separately, but when the electric field turns on, is affected by a quantity proportional to the current density from the case with $\mathcal{E}$ null. Likewise, while the magnetic field is applied only, the function $j_{y}(x)$ has contributions of the components mixed and when the electric field appears, the starting current density is affected by an amount proportional to $\rho_{\alpha}(x)$ from the case with $\mathcal{E}=0$ (see Eq.~(\ref{density}) and Fig.~\ref{fig:rhojy}). Similar results are found for the eigenstates $\Psi_{n}(x,y)$ (see Eq.~(\ref{density_2}) and Fig.~\ref{fig:rhoN}).

As was mentioned before, the main role of the coherent states in quantum mechanics is, in general, to describe the behavior of a system in a semi-classical approach, although that aim is not easy for all quantum systems. In our case, when Dirac electrons interact with an external magnetic field only, for a phase variation of the eigenvalue $\alpha$ that characterizes the coherent state, the probability density does not change its shape: a Gaussian distribution on the $x$-axis. In the added electric field case, the probability density is accumulated in some particular points along the $x$-direction as $\varphi$ changes. This means that is more probable to find electrons in certain spatial regions choosing the correct eigenvalue phase. Likewise, the current density changes as the parameter $\beta$ increases, taking positive values in the points where $\vert\Psi_{\alpha}\vert^2$ is longer on $x$-axis. Thus, we can assume that in such points electrons not only reduce their group velocity, but also their motion direction changes as a result of the presence of a constant electric field directed along the positive $x$-axis (Fig.~\ref{fig:rhojy}). From a semi-classical point of view, an electron in an external magnetic field describes a circular orbit, which is equivalent to being able to detect the particle in wherever place on the trajectory. When an electric field is applied, the charged carriers move in a spiral way giving place to points in which their speed decreases and they can be detected with larger probability (Fig.~\ref{fig:graphene}). Furthermore, as the electric field strength increases, such points appear more to the left of the $x$-axis each time, due to the attractive forces generated by the electric field.

\begin{figure}[!ht]
	\centering
	\includegraphics[width=0.8\textwidth]{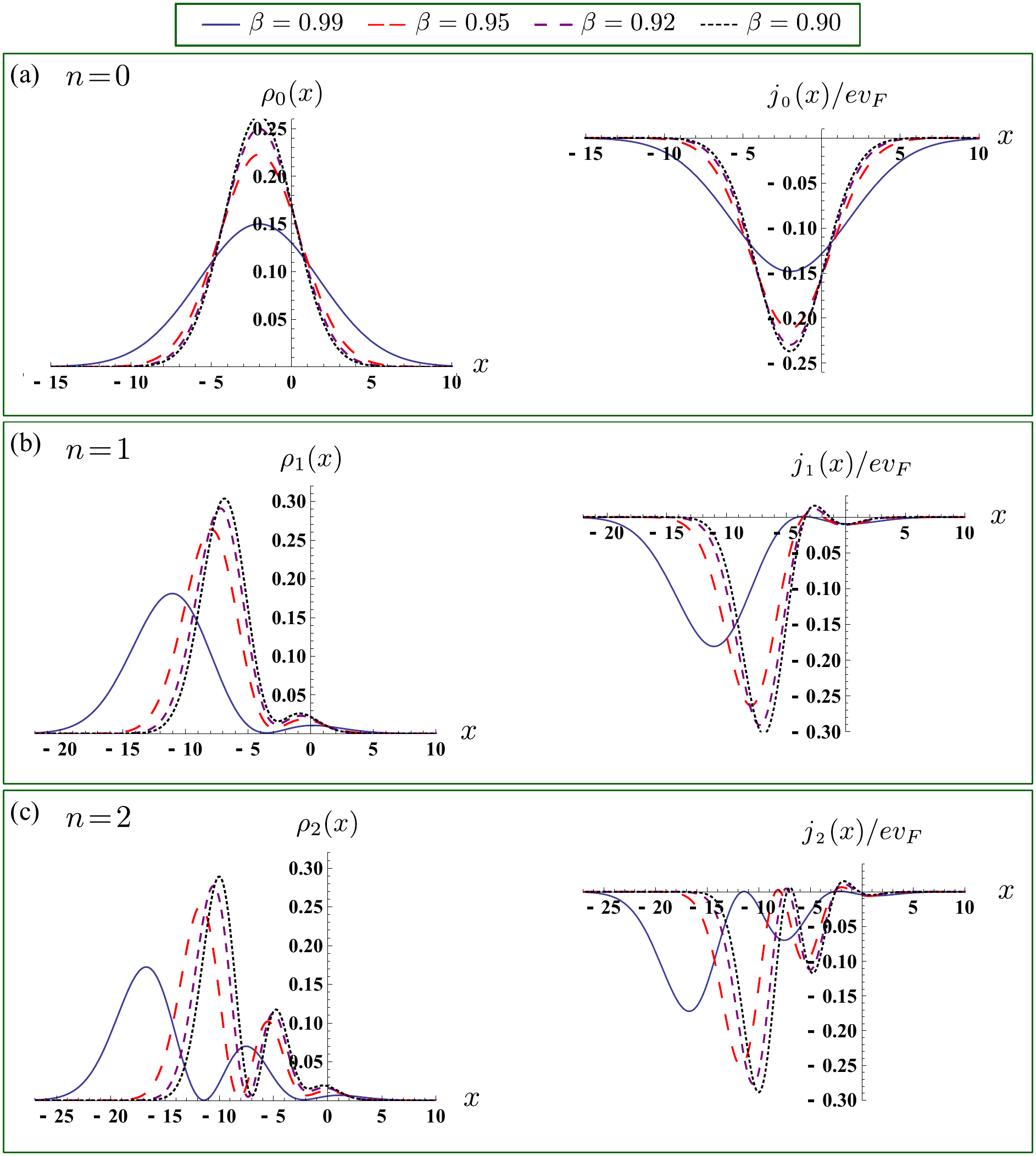}
	\caption{\label{fig:rhoN_1}Upper panels show the probability density $\rho_n(x)$ (left-hand) and the $y$-current density $j_{n}(x)/e\,v_{\rm F}$ (right-hand) for three different Landau levels $n=0,1,2$ and values of $\beta$ close to $1$. In all these cases $B=1/2$, $k=\omega_{\rm B}=1$.}
\end{figure}

\begin{figure}[!ht]
	\centering
	\includegraphics[width=0.45\textwidth]{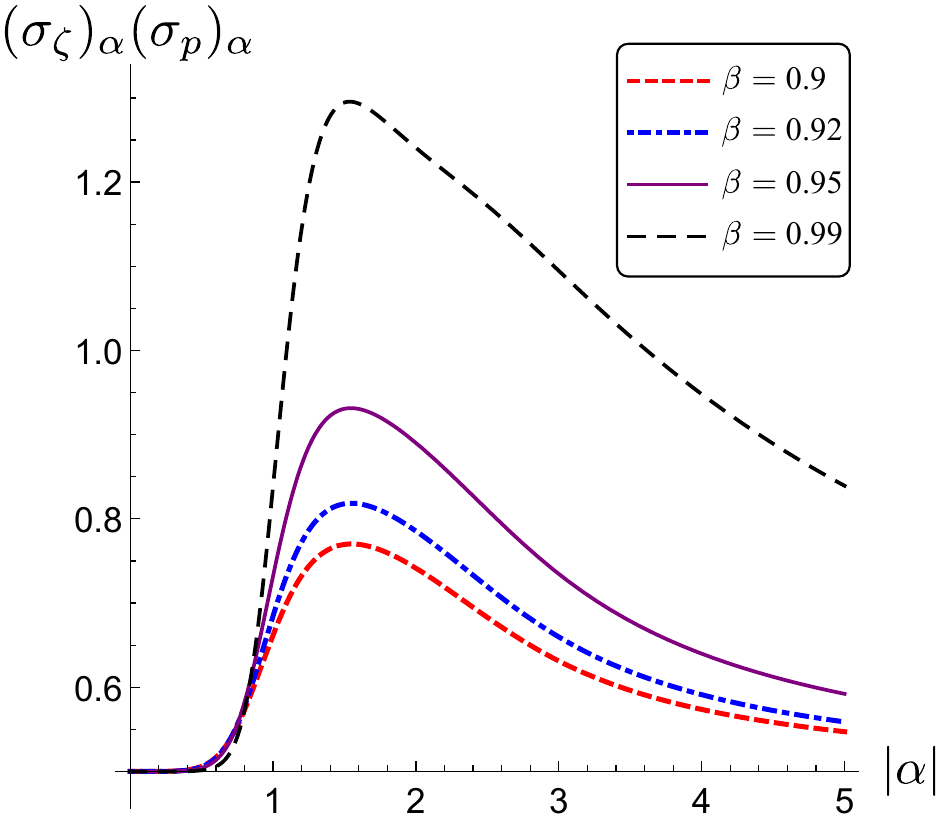}
	\caption{\label{fig:HUR_1}HUR for the coherent states $\Psi_{\alpha}(x,y)$ as function of $\vert\alpha\vert$ for values of $\beta$ close to $1$.}
\end{figure}

Besides, following \cite{huang18}, the average velocity in $y$-direction is calculated as
\begin{equation}
\langle v_{y}\rangle_{\alpha}=\frac1\hbar\frac{\partial}{\partial k}\langle H\rangle_{\alpha}=\left[2\exp\left(\vert\tilde{\alpha}\vert^2\right)-1-2\beta\Re\left(\tilde{\alpha}\right)\sum_{n=0}^{\infty}\frac{\vert\tilde{\alpha}\vert^{2n}}{n!\sqrt{n+1}}\right]^{-1}\left[v_{\rm d}\left(1-2\exp\left(\vert\tilde{\alpha}\vert^2\right)\right)\right].
\end{equation}
As we can see in Fig.~\ref{fig:vy}, as $\vert\alpha\vert$ grows up, the average velocity $\vert\langle v_{y}\rangle_{\alpha}\vert$ tends to a smaller (bigger) constant value than $v_{\rm d}$ for $\pi/2\leq\varphi\leq3\pi/2$ ($-\pi/4<\varphi<\pi/4$).  The minus sign in $\langle v_{y}\rangle_{\alpha}$ indicates the velocity is directed along the negative $y$-axis.


\subsubsection*{Collapse of Landau levels ($\beta\rightarrow1$)}
Figure~\ref{fig:rhoN_1} shows the behavior of the probability and current densities along the $x$-axis for the states in Eq.~(\ref{20}) when $\beta$ takes values close to $1$. We can see that while the sign of the function $j_{y}(x)$ holds negative, the amplitude of the maximum probability for such values is less than when $\beta<1$ (Fig.~\ref{fig:rhoN}), indicating that when $\mathcal{E}\approx v_{\rm F}B/c$, the probability of finding a Dirac particle on the left side of the $x$-axis decreases.

On the other hand, as the drift velocity $v_{\rm d}$ approximates to Fermi velocity $v_{\rm F}$, the Heisenberg uncertainty relation is longer than $1/2$ only for real values of $\alpha$ (see Figs.~\ref{fig:HUR} and~\ref{fig:HUR_1}), for which the position-like variance $\sigma_{\zeta}$ remains always larger than the momentum-like one $\sigma_{p}$. Additionally, it is worth to mention that for $\alpha\in\mathbb{C}$ for which the HUR is not equal to $1/2$, the mean energy $\langle H\rangle_{\alpha}$ is tilted as also $\beta$ increases  (Fig~\ref{fig:Hbeta}), indicating that such values would correspond to a larger mean energy. 

In summary, since Landau levels already collapse in the limit value $\mathcal{E}_{c}=v_{\rm F}B/c$, our results are not valid close to that critical point due to the bounded states would disappear becoming into scattering states and we would have that $\vert\langle v_{y}\rangle_{\alpha}\vert>v_{\rm d}\approx v_{\rm F}$ \cite{knn09}. Actually, in~\cite{cheng13} the critical value $\beta=1$ is analyzed showing that existence of no bound states, while for $\beta>1$, {\it i.e.} a weak magnetic field, the problem holds solvable and leads to a bit different spectrum behavior~\cite{nathroy14}.

\section{Concluding remarks}\label{conclusions}
We have solved the two-dimensional Dirac-Weyl equation with an external electric field by using a clear algebraic method that prevents us to implement techniques of special relativity. Also, we have been able to obtain the known results of the system interacting only with a magnetic field by taking the limit $\beta\rightarrow0$.

Likewise, we have build the coherent states for a graphene sample on $xy$-plane interacting with both electric and magnetic fields. In addition, our results are not only in agreement with those of classical mechanics but also allow to establish a model that describes the effects of both fields on charge carriers in a stationary regime through physical quantities such as probability and current densities, Heisenberg uncertainty relation and mean energy. Our main conclusion is that the electric field modifies the location of the quasi-particles in graphene. This semi-classical description could be implemented to analyze electronic or transport properties in graphene or other 2D DMs, as occurs for the so-called Bloch oscillations~\cite{huang18}.


\section*{Acknowledgments}
The authors acknowledge V\'{i}t Jakubsk\'{y} and the anonymous referee for their valuable comments to improve this work. M.C.C. acknowledges the CONACyT fellowship 301117. M.O.L. acknowledges to CONACyT for a postdoctoral fellowship.

\appendix
\numberwithin{equation}{section}
\section{Eigenvalues and eigenstates}\label{eigenstates}
In order to find the solutions of our problem, we proceed as~\cite{o14}. Hence, multiplying by $-i\sigma_{x}$ to the left of Eq.~(\ref{8}), we get:
\begin{equation}\label{95}
\left[\frac{1}{l_{\rm B}}\frac{d}{d\xi}\mathbb{I}_{2}-i\left(\epsilon_0-\frac{\beta \xi}{l_{\rm B}}\right)\sigma_{x}\mathbb{I}_{2}+i\frac{\xi}{l_{\rm B}}\sigma_{x}\sigma_{y}\right]\Psi(\xi)=0,
\end{equation}
where $\epsilon_0=E/\hbar v_{\rm F} +k\beta$.
Differentiating this equation with respect to $\xi$ and manipulating a bit, we obtain the following equation
\begin{equation}
\left[\left(\frac{d^2}{d\xi^2}+\left(\epsilon_0l_{\rm B}-\beta \xi\right)^2-\xi^2\right)\mathbb{I}_{2}+i\left(\sigma_{x}\beta+\sigma_{x}\sigma_{y}\right)\right]\Psi(\xi)=0,
\end{equation}
whose solutions can be expressed as $\Psi(\xi)_{\lambda}=\psi_{\lambda}(\xi)\chi_{\lambda}$, where $\chi_{\lambda}$ is an eigenvector of the complex symmetric matrix $\mathbb{K}=i\left(\sigma_{x}\beta+\sigma_{x}\sigma_{y}\right)$ with eigenvalue $\lambda$ and $\psi_{\lambda}(\xi)$ is a scalar function that satisfies the differential equation
\begin{equation}\label{98}
\left[\frac{d^2}{d\xi^2}+\left(\epsilon_0l_{\rm B}-\beta \xi\right)^2-\xi^2+\lambda\right]\psi_{\lambda}(\xi)=0,
\end{equation}
In order to simplify the above equation, the variable $\zeta$ is defined as
\begin{equation}
\zeta=\xi(1-\beta^2)^{1/4}+\frac{\beta\epsilon_0l_{\rm B}}{(1-\beta^2)^{3/4}}=\frac{(1-\beta^2)^{1/4}}{l_{\rm B}}\left[x+l_{\rm B}^2k+\frac{\beta\epsilon_0l_{\rm B}^2}{1-\beta^2}\right],
\end{equation}
where $\beta$ must fulfill the condition $0\leq\beta<1$ for keeping real values of $\zeta$ and prevent indeterminacy. Likewise, this implies that $v_{\rm d}=\mathcal{E}c/B<v_{\rm F}$, in contrast with the classical drift velocity, in which there is no restriction for the electric field strength. Hence, we obtain the Weber equation
\begin{equation}\label{100}
\left[\frac{d^2}{d\zeta^2}-\zeta^2+\frac{\epsilon_0^2l_{\rm B}^2}{(1-\beta^2)^{3/2}}+\frac{\lambda}{(1-\beta^2)^{1/2}}\right]\psi_{\lambda}(\zeta)=0,
\end{equation}
where the dimensionless potential $V_{\beta}(\zeta)$ can be identified as
\begin{equation}
V_{\beta}(\zeta)=\zeta^2-\frac{\lambda}{(1-\beta^2)^{1/2}}.
\end{equation}

On the other hand, the eigenvalues $\lambda$ of the matrix $\mathbb{K}$ turn out to be $\sigma(\mathbb{K})=\{\lambda_k=(-1)^{k}(1-\beta^2)^{1/2}\}$ with $k=1,2$, while the corresponding normalized eigenvectors are given by
\begin{align}
S &=\left\{
\chi_{\lambda_1}=\frac{1}{\sqrt{2}}\left(\begin{array}{c}
\sqrt{C_+} \\
-i\sqrt{C_-}
\end{array}\right), \quad \chi_{\lambda_2}=\frac{1}{\sqrt{2}}\left(\begin{array}{c}
-\sqrt{C_-} \\
i\sqrt{C_+} \\
\end{array}\right)\right\}, 
\end{align}
where $C_{\pm}=1\pm(1-\beta^2)^{1/2}$. Substituting the eigenvalues $\lambda_k$ in Eq.~(\ref{100}) and taking $\psi_{\lambda}(\zeta)=\exp\left(-\zeta^2/2\right)f_{\lambda}(\zeta)$, one gets the following ODE:
\begin{equation}
f''_{\lambda}(\zeta)-2\zeta f'_{\lambda}(\zeta)+\left(\frac{\epsilon_0^2l_{\rm B}^2}{(1-\beta^2)^{3/2}}-1+(-1)^{k}\right)f_{\lambda}(\zeta)=0, \quad k=1,2.
\end{equation}

From here, the results shown in Eqs.~(\ref{energy})-(\ref{106}) are followed. Besides, the procedure implemented to solve (\ref{8}) has also been successfully applied for reproducing some results in \cite{mht09}.

\section{Matrix $\mathbb{M}$}\label{matrixM}
In this part, we focus on describing some features of matrix $\mathbb{M}$ in Eq.~(\ref{11}).

Since $\mathbb{M}$ is a hermitian matrix, it can be diagonalized, {\it i.e.}, there is a diagonal matrix $\mathbb{D}$ and a unitary matrix $\mathbb{U}$ such that $\mathbb{M}=\mathbb{U}\mathbb{D}\mathbb{U}^{-1}$.

After finding the $\mathbb{M}$-eigenvalues,
\begin{equation}
\sigma(\mathbb{M})=\left\{\mu_{k}=\frac{\sqrt{C_{+}}+(-1)^k\sqrt{C_{-}}}{\sqrt{2}}\right\}, \quad k=1,2,
\end{equation}
and the corresponding eigenvectors,
\begin{equation}
S=\left\{\chi_{\mu_1}=\frac{1}{\sqrt{2}}\left(\begin{array}{c}
1 \\
i
\end{array}\right), \quad \chi_{\mu_2}=\frac{1}{\sqrt{2}}\left(\begin{array}{c}
i \\
1
\end{array}\right)\right\},
\end{equation}
it follows that matrix $\mathbb{M}$ can be expressed as
\begin{equation}
\mathbb{M}=\mathbb{U}(\pi/4)\mathbb{D}\mathbb{U}^{-1}(\pi/4),
\end{equation}
where $\mathbb{D}={\rm diag}\left(\mu_1,\,\mu_2\right)$ and $\mathbb{U}(\tau)=\exp\left(i\tau\sigma_{x}\right)$. Therefore, $\mathbb{M}$ is obtained from rotations about the $x$-axis.

\section{Probability and current densities}\label{densities}
We can obtain the expressions for  the probability density $\rho_{n}(x)=\vert\Psi_{n}(x)\vert^2$ and the current density $j_{x/y}(x)=ev_{\rm F}\Psi_{n}^{\dagger}(x)\sigma_{x/y}\Psi_{n}(x)$ in the $x/y$-direction by identifying the following relations:
\begin{subequations}
	\begin{align}
\mathbb{M}^{\dagger}\mathbb{M}&=\mathbb{I}_2-\beta\sigma_{y}, \\
\mathbb{M}^{\dagger}\sigma_{x}\mathbb{M}&=\sqrt{1-\beta^2}\,\sigma_{x}, \\
\mathbb{M}^{\dagger}\sigma_{y}\mathbb{M}&=\sigma_{y}-\beta\mathbb{I}_2.
\end{align}
\end{subequations}

Thus, the probability density $\rho_{n}(x)$ and the current densities $j_{x/y}(x)$ turn out to be, respectively: 
\begin{subequations}\label{density_2}
	\begin{align}
	\rho_n(x)&= \vert\bar{\Phi}_{n}(x,y)\vert^2-\beta\bar{\Phi}_{n}^{\dagger}(x,y)\sigma_{y}\bar{\Phi}_{n}(x,y), \\
	j_{x}(x)&=0, \\
	j_{y}(x)&=ev_{\rm F}\bar{\Phi}_{n}^{\dagger}(x,y)\sigma_{y}\bar{\Phi}_{n}(x,y)-ev_{\rm d} \vert\bar{\Phi}_{n}(x,y)\vert^2.
	\end{align}
\end{subequations}

Besides, if $s$ represents a scalar operator associated to any observable quantity that has been promoted to a matrix one through the relation $s\rightarrow\mathbb{S}=s\otimes\mathbb{I}$, we have that
\begin{equation}\label{32}
\langle\mathbb{S}\rangle=\langle\Psi_{n}\vert\mathbb{S}\vert\Psi_{n}\rangle=\langle\Phi_{n}\vert\mathbb{S}\vert\Phi_{n}\rangle-\beta\langle\Phi_{n}\vert s\,\sigma_{y}\vert\Phi_{n}\rangle.
\end{equation}

\section{Matrix operators $\Theta^{\pm}$}\label{ladderop}
	The action of the matrix operators defined in Eq.~(\ref{ladder}) on the eigenstates $\bar{\Phi}_{n}(x,y)$ is given by:
	\begin{subequations}
		\begin{align}
		\Theta^{-}\bar{\Phi}_{n}(x,y)&\equiv\Theta^{-}\bar{\Phi}_{n}(\zeta_{n},y)=\frac{\exp\left(i\,\delta\right)}{\sqrt{2^{\delta_{1n}}}}\sqrt{n}\bar{\Phi}_{n-1}(\zeta_{n-1},y), \quad n=0,1,2,\dots,\\ \Theta^{+}\bar{\Phi}_{n}(x,y)&\equiv\Theta^{-}\bar{\Phi}_{n}(\zeta_{n},y)=\exp\left(-i\,\delta\right)\sqrt{n+1}\bar{\Phi}_{n+1}(\zeta_{n+1},y), \quad n=1,2,3,\dots.
		\end{align}
	\end{subequations}
	According to these relations, the operator $\Theta^{+}$ cannot be considered as a creation operator since its action on the state $\bar{\Phi}_{0}$ does not generate other eigenstate of the Hilbert space $\mathcal{H}$. As a consequence, the commutation relation $[\Theta^{-},\Theta^{+}]=\mathbb{I}_{2}$ only fulfills for $n\geq2$.
	
	On the other hand, if we consider the operator
	\begin{equation}
	\tilde{\Theta}^{+}=\exp(-i\delta)\sum_{n=0}\left(\begin{array}{c c}
	\theta_{n}^{+}\frac{\sqrt{N+2}}{\sqrt{N+1}} & -i\eta\sqrt{N+1} \\
	i\eta(\theta_{n}^{+})^2\frac{1}{\sqrt{N+1}} & \theta_{n}^{+}
	\end{array}\right)\mathcal{P}(n)\mathcal{T}^{+},
	\end{equation}
	such that
	\begin{equation}
	\tilde{\Theta}^{+}\bar{\Phi}_{n}(x,y)\equiv\tilde{\Theta}^{+}\bar{\Phi}_{n}(\zeta_{n},y)=\sqrt{2^{(2-\delta_{0n})}}\exp(-i\delta)\sqrt{n+1}\bar{\Phi}_{n+1}(\zeta_{n+1},y), \quad n=0,1,2,\dots,
	\end{equation}
	we would be able to obtain excited states from the fundamental $\bar{\Phi}_{0}$ as follows:
	\begin{equation}\label{eigenstate}
	\bar{\Phi}_{k}(x,y)\equiv\bar{\Phi}_{k}(\zeta_{k},y)=\frac{\exp\left(ik\delta\right)}{\sqrt{2^{(2k-1)}k!}}(\tilde{\Theta}^+)^{k}\bar{\Phi}_{0}(\zeta_{0},y), \quad k=1,2,3,\dots.
	\end{equation}
	This fact shows that, although $\tilde{\Theta}^{+}$ is not the adjoint of $\Theta^{-}$, it works as a creation operator. Furthermore, it is straightforward to verify that
	\begin{equation}
	[\Theta^{-},\tilde{\Theta}^{+}]\bar{\Phi}_{n}=c(n)\bar{\Phi}_{n}, \quad c(n)=\begin{cases}
	1, & n=0,\\
	3, & n=1,\\
	2, & {\rm otherwise}.
	\end{cases}
	\end{equation}
	Hence, these results suggest that, up to a constant factor, $\tilde{\Theta}^{+}$ and $\Theta^{-}$ would be linked to the so-called $\mathcal{D}$ pseudo-bosonic operators~\cite{trifonov09,bagarello13,bagarello132,bagarello15,bagarello17} that arise by modifying the canonical commutation relation $[c,c^{\dagger}]=1$, which is replaced with a similar commutation rule, namely, $[a,b]=1$ where $b\neq a^{\dagger}$. This discussion takes relevance when, for instance, the construction of coherent states is performed by the Perelomov group theoretical approach~\cite{perelomov72}, in which the algebraic structure of the ladder operators is required.

\section{Heisenberg uncertainty relation}\label{heisenberg}
Using the states in Eq.~(\ref{40}) and according to (\ref{32}), the mean values of the operators $\mathbb{S}_q$ and $\mathbb{S}^2_q$ are, respectively (see Fig.~\ref{fig:HUR}):
\small
\begin{subequations}
	\begin{align}
	\nonumber\langle\mathbb{S}_q\rangle_{\alpha}&=\frac{1}{\sqrt{2}i^q}\left[2\exp\left(\vert\tilde{\alpha}\vert^2\right)-1-2\beta\Re\left(\tilde{\alpha}\right)\sum_{n=0}^{\infty}\frac{\vert\tilde{\alpha}\vert^{2n}}{n!\sqrt{n+1}}\right]^{-1}\left[\left(\tilde{\alpha}+(-1)^q\tilde{\alpha}^\ast\right)\Bigg(\exp\left(\vert\tilde{\alpha}\vert^2\right)+\right.\\
	&\quad\left.\sum_{n=1}^{\infty}\frac{\vert\tilde{\alpha}\vert^{2n}}{\sqrt{(n-1)!(n+1)!}}\Bigg)-\beta\left((\tilde{\alpha}^2+(-1)^q\tilde{\alpha}^{\ast 2})\sum_{n=0}^{\infty}\frac{\vert\tilde{\alpha}\vert^{2n}}{n!\sqrt{n+2}}+(1+(-1)^q)\sum_{n=1}^{\infty}\frac{\vert\tilde{\alpha}\vert^{2n}}{n!}\sqrt{n}\right)\right], \\
	\nonumber\langle\mathbb{S}^2_q\rangle_{\alpha}&=\frac{1}{2}\left[2\exp\left(\vert\tilde{\alpha}\vert^2\right)-1-2\beta\Re\left(\tilde{\alpha}\right)\sum_{n=0}^{\infty}\frac{\vert\tilde{\alpha}\vert^{2n}}{n!\sqrt{n+1}}\right]^{-1}\Bigg\{1+4\vert\tilde{\alpha}\vert^2\exp\left(\vert\tilde{\alpha}\vert^2\right)+(-1)^q(\tilde{\alpha}^2+\tilde{\alpha}^{\ast 2})\\
	&\nonumber\quad\times\Bigg(\exp\left(\vert\tilde{\alpha}\vert^2\right)+\sum_{n=1}^{\infty}\frac{\sqrt{n+1}\,\vert\tilde{\alpha}\vert^{2n}}{\sqrt{(n-1)!(n+2)!}}\Bigg)-\beta\Bigg[(\tilde{\alpha}+\tilde{\alpha}^{\ast})\left(\sum_{n=0}^{\infty}\frac{\vert\tilde{\alpha}\vert^{2n}}{n!\sqrt{n+1}}(2n+1)\right.\\
	&\quad\left.+(-1)^q\sum_{n=1}^{\infty}\frac{\vert\tilde{\alpha}\vert^{2n}}{\sqrt{n!(n-1)!}}\right)+(-1)^q(\tilde{\alpha}^3+\tilde{\alpha}^{\ast 3})\sum_{n=0}^{\infty}\frac{\vert\tilde{\alpha}\vert^{2n}}{n!\sqrt{n+3}}\Bigg]\Bigg\}.
	\end{align}
\end{subequations}
\normalsize

\section{Completeness relation}
Let us consider the  Hilbert space $\mathcal{H}$ spanned by the DW eigenstates, $\mathcal{H}={\rm span }\{\bar{\Phi}_n\vert n=0,1,2,\dots\}$, which fulfill the completeness relation
\begin{equation}\label{cr}
\ket{\bar\Phi_0}\bra{\bar\Phi_0}+\sum_{n=1}^{\infty}\ket{\bar\Phi_n}\bra{\bar\Phi_n}\equiv\mathbb{I}_{2}.
\end{equation}

By defining $r=\vert\alpha\vert^2$, we take the measure as:
\begin{equation}
d\mu(\alpha)=\frac{2\exp\left(r^2\right)-1}{2\pi}r\exp\left(r^2\right)drd\theta,
\end{equation}
\noindent       
we obtain the overcompleteness relation of the coherent states as follows:
\small
\begin{align}
\nonumber \frac{\ket{\bar\Phi_0}\bra{\bar\Phi_0}}{2}\,+&\int_{\mathbb
C}d\mu({\alpha})\ket{\bar\Phi_n}\bra{\bar\Phi_n}\\
\nonumber =&\quad\frac{\ket{\bar\Phi_0}\bra{\bar\Phi_0}}{2}+\int_{\mathbb
C}\frac{d\mu({\alpha})}{2\exp\left(r^2\right)-1}\left[\ket{\bar\Phi_0}+\sum_{n=1}^{\infty}\frac{\sqrt{2}\tilde{\alpha}^n}{n!}\ket{\bar\Phi_n}\right]\left[\bra{\bar\Phi_0}+\sum_{m=1}^{\infty}\frac{\sqrt{2}\tilde{\alpha}^m}{m!}\bra{\bar\Phi_m}\right]\\
\nonumber =&\quad\frac{\ket{\bar\Phi_0}\bra{\bar\Phi_0}}{2}+\frac{1}{2\pi}\int_{0}^{\infty}\int_{0}^{2\pi}\bigg[\ket{\bar\Phi_0}\bra{\bar\Phi_0}+\sum_{n=1}^{\infty}\frac{\sqrt{2}}{\sqrt{n!}}\ket{\bar\Phi_n}\bra{\bar\Phi_0}r^ne^{in\theta}\\
&+\sum_{m=1}^{\infty}\frac{\sqrt{2}}{\sqrt{m!}}\ket{\bar\Phi_0}\bra{\bar\Phi_m}r^me^{-im\theta}+\sum_{n,m=1}^{\infty}\frac{2}{\sqrt{n!m!}}\ket{\bar\Phi_n}\bra{\bar\Phi_m}r^{n+m}e^{i(n-m)\theta}
\bigg]r\exp\left(r^2\right)drd\theta.
\end{align}
\normalsize

Taking the variable change $t=r^2$ and integrating over $\theta$ and $t$, we get finally:
\begin{equation}
\frac{\ket{\bar\Phi_0}\bra{\bar\Phi_0}}{2}+\int_{\mathbb
C}d\mu({\alpha})\ket{\bar\Phi_n}\bra{\bar\Phi_n}=\ket{\bar\Phi_0}\bra{\bar\Phi_0}+\sum_{n=1}^{\infty}\frac{\ket{\bar\Phi_n}\bra{\bar\Phi_n}}{n!}\Gamma(n+1),
\end{equation}
which reduces to Eq.~(\ref{cr}). 

\bibliographystyle{ieeetr}
\bibliography{biblio}

\end{document}